# Accurate Estimation of Transport Coefficients Using Model-free Time Correlation Functions in Equilibrium Simulations

Xin Liu,[1] Xuhong Guo,[1,*] and Qi Liao[2,†]

[1] State Key Laboratory of Chemical Engineering,
East China University of Science and Technology, Shanghai 200237, China

[2] Institute of Chemistry, Chinese Academy of Sciences, Beijing 100190, China

Transport coefficients, such as the diffusion coefficient and shear viscosity, are important material properties that are calculated in computer simulations. In this study, the criterion for the best estimation of viscosity, as an example of transport coefficients, is determined by using the Green-Kubo formula without any artificial models. The related algorithm is given by the estimation of the viscosities of polyethylene oxide solutions by using a molecular dynamics simulation for testing. The algorithm can be used in the simulations of complex systems with a long tail of correlations typically found in macromolecular and biological simulation systems.

***Introduction.*** – Transport coefficients can be defined as the response of a system to a perturbation and can be related to the infinite time integral of an equilibrium time correlation function [1]. As an important transport coefficient of materials, there have been several methods for computing the shear viscosity $\eta$ of fluids from molecular dynamics (MD) simulation: (1) simulating systems in equilibrium based on pressure or momentum fluctuations; (2) applying a perturbation, such as an extra force or Couette flow, and driving systems out of equilibrium and measuring their response [1-3]. Equilibrium molecular dynamics methods (EMDs) are superior to non-equilibrium molecular dynamics methods (NEMDs) in various ways. For example, they can be used in smaller simulation system sizes without compromising the accuracy of results [4], and owing to their conciseness, the Green-Kubo formula, which involves the integrating of the stress relaxation shear modulus $G(\tau)$, is commonly used to estimate transport coefficients:

$$\eta = \int_0^\infty G(\tau)d\tau, \qquad (1)$$

$$G(\tau) = \frac{V}{k_B T} \langle P_{\alpha\beta}(\tau)P_{\alpha\beta}(0)\rangle, \qquad (2)$$

where $V$ and $T$ are the volume and temperature of the simulation system, respectively, $k_B$ is the Boltzmann constant, $P_{\alpha\beta}$ refers to the off-diagonal components of the pressure tensor, and the angle bracket is a canonical ensemble average. Theoretically, $G(\tau)$ can decay to zero as time tends to infinity; hence, the integral in Eq. (1) can also achieve a constant value of $\eta$. In practical research, however, due to the significant fluctuation of the pressure in the simulation system, we must choose an approximate integration time $t_{cutoff}$, which is long enough such that $G(\tau)$ can achieve a statistically zero region but short enough that the statistical noise does not occupy a dominant contribution, to yield a reliable result [5].

In early research, the Green-Kubo formula was widely used to calculate the shear viscosity of Lennard-Jones fluids [6], atomic liquids such as liquid iron [7], liquid aluminum [8], and other liquid metals [9] with significant application value, in addition to some simple small molecular liquids, especially water [10]. Throughout the associated research reported in the literature, two different procedures are included: (1) Some researchers determine a $t_{cutoff}$ using an arbitrary method [7-9]. For example, Zhang et al. [7]

assume that the time when the decay curve of $G(t)$ first reaches zero or when the integral curve of $\eta(t)$ initially arrives at the maximum value, is an effective $t_{cutoff}$, the upper limit of the Green-Kubo formula integral. (2) Furthermore, some researchers selected a plateau region of the Green-Kubo formula integral by visual inspection and then considered its average value as the final estimated $\eta$ [11]. In some cases, especially relatively-high viscosity fluids, however, diverse results may be obtained using different procedures with the same system because none of them truly minimize the statistical error [5,12].

To improve the statistical noise and extend the applicability of the Green-Kubo formula, a few researchers have proposed that the decay curve of $G(t)$ or the integral curve of $\eta(t)$ may be fitted using an analytic function and model, yielding a robust $\eta$ value. In 2015, Maginn et al. [13] proposed a time decomposition method that fits the Green-Kubo formula with a double-exponential function. Subsequently, this method was successfully used to calculate $\eta$ of relatively high viscosity liquids, such as ionic liquids [14], hydrocarbons under high temperature and high pressure [15-17], and some multicomponent mixtures [18]. Furthermore, there are several different types of semiempirical analytic forms for $G(t)$ or $\eta(t)$ reported in the literature [19], such as a double-exponential function, a gaussian function and other more complicated function forms [20]. To determine the fitting parameters, the real theoretical models of their simulation system must be known and some assumptions must be made; further, the biggest advantage of these analytic forms could be that the simulation results are combined with the physical models. Although the research on the Green-Kubo formula has yielded many remarkable outcomes, each procedure is based on unique assumptions and only adapts to specific systems [21-24]. In other words, computing a dependable $\eta$ value based on the Green-Kubo formula remains challenging for typical complex systems, such as the macromolecular and biological systems, whose theoretical models are unknown.

In this Letter, we proposed a criterion for the best estimation of $\eta$ using the Green-Kubo formula through EMD simulation and without employing any anticipated theoretical modes. Our algorithm, in contrast to all prior studies [21,22], just uses pertinent data obtained during the EMD simulation and does not rely on any premise or assumptions. In the EMD simulation, the measured viscosity at time $t$ includes the correlation tail and error:

$$\eta(t) = \int_0^t [G(\tau) + \Delta G(\tau)]d\tau \qquad (3a)$$

$$= \eta - \int_t^\infty G(\tau)d\tau + \int_0^t \Delta G(\tau)d\tau. \qquad (3b)$$

where $\int_t^\infty G(\tau)d\tau$ is the tail of the integration of Eq. (1), and $\int_0^t \Delta G(\tau)d\tau$ is the error of the simulation in Eq. (2). Thus, the total error of shear viscosity using Eq. (1) and Eq. (2) from simulation data may be estimated through the following:

$$\Delta\eta(t) = \eta - \eta(t) \qquad (4a)$$

$$= \int_t^\infty G(\tau)d\tau - \int_0^t \Delta G(\tau)d\tau \qquad (4b)$$

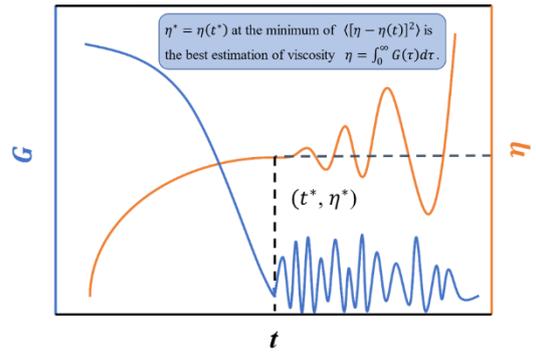

FIG.1. Criterion for the best estimation of cutoff time of integration of the Green-Kubo formula. Owing to the long tail of the Green-Kubo formula and the noise of the simulation data, it is difficult to derive a reliable viscosity.

The best estimation of the integral time $t^*$ can be determined by the minimum of the errors of two contributions, as shown in Fig. 1.

The minimum of the errors can be estimated as follows:

$$\langle[\Delta\eta(t)]^2\rangle = \left(\int_t^\infty G(\tau)d\tau\right)^2$$
$$+ \left\langle\left(\int_0^t \Delta G(\tau)d\tau\right)^2\right\rangle. \quad (5)$$

Because $\langle\Delta G(\tau)\rangle = 0$ and $\langle\Delta G(\tau)\cdot\Delta G(\tau+t)\rangle = \Delta G^2(t)\delta(t)$, the differentiation of Eq. (5) gives

$$\frac{d\langle[\Delta\eta(t)]^2\rangle}{dt} = -2G(t)\int_t^\infty G(\tau)d\tau$$
$$+ 2\left\langle\Delta G(t)\left(\int_0^t \Delta G(\tau)d\tau\right)\right\rangle \quad (6a)$$

$$= -2G(t)\int_t^\infty G(\tau)d\tau$$
$$+ 2\Delta G^2(t)\delta(t). \quad (6b)$$

Thus, the final result of the best cutoff time of integral $t^*$ is given by the equation

$$G(t^*)\int_{t^*}^\infty G(\tau)d\tau = \langle\Delta G^2(t^*)\rangle\delta(t^*). \quad (7a)$$

$$G(t^*)\left[\eta - \int_0^{t^*} G(\tau)d\tau\right] = \langle\Delta G^2(t^*)\rangle\delta(t^*). \quad (7b)$$

In practice, Eq. (7b) is calculated based on the MD simulation data and compared iteratively to obtain the cutoff time $t^*$ and viscosity $\eta(t^*)$ using the following definitions

$$E_1(t) = G(t)[\eta(t^*) - \eta(t)] \quad (8)$$
$$E_2(t) = \langle\Delta G^2(t)\rangle. \quad (9)$$

***Test vehicle.***—As a test vehicle, our algorithm was applied to calculate the viscosity $\eta$ of dilute polyethylene oxide (PEO) aqueous solution across different concentrations. Initially, a reasonable initial simulation model (see Fig. 2) was successfully constructed using Enhanced Monte Carlo (EMC) [25] based on optimized potentials for liquid simulations - united atom (OPLS-UA) force field [26]. Considering previous reports [15,27] with respect to the system size effects on $\eta$, greater system sizes do not have a significant impact on the $\eta$ value calculated using EMD simulation. Therefore, a cubic simulation box with 10,000 atoms and periodic boundary conditions (PBC) were used

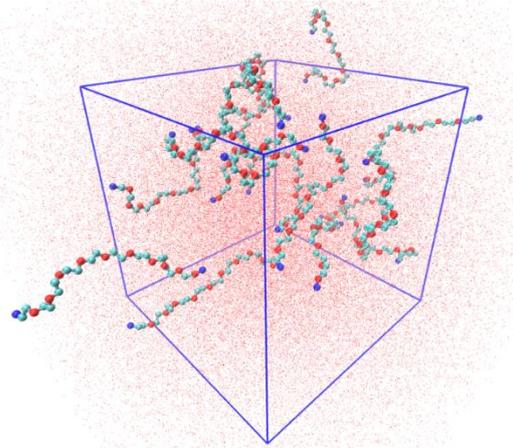

FIG. 2. Snapshot of PEO solution with weight fraction 8%. For each PEO chain with $X - [CH_2 - CH_2 - O]_n - Y$ chemical formula, the number of monomers $n$ is 10, and the terminal group is $X = Y = CH_3$. For clarity, the solvent water molecules are shown with red dots. The carbon and oxygen in the monomers and the carbon in the terminal groups are shown in cerulean, red, and blue, respectively.

for all EMD simulations with Large-scale Atomic/Molecular Massively Parallel Simulator (LAMMPS) package [28]. For all intermolecular interactions, a real space cutoff distance of 9.5 Å was employed. Long-range electrostatic interactions were calculated using the particle-particle particle-mesh (PPPM) method [29]. The velocity Verlet method was used to integrate the equations of motion with a timestep of 2 fs. The isothermal-isobaric (NPT) ensemble was used to maintain the temperature at 300 K and the pressure at 1 atm. Simulations were performed at seven different weight densities of PEO (0, 4%, 8%, 10%, 12%, 16%, and 20%) with a time of 10 ns and longer. The "fix ave/correlate/long" was used efficiently calculate time correlation functions on-the-fly over extremely long-time windows. To ensure that our simulation system is in the dilute solution range, we recorded atomic coordinates every 1000 fs in the EMD simulation with a total run time of 1 ns. Next, we computed the end-to-end distance $R_e$ with relatively high concentration. Furthermore, the overlap concentration $C^*$, indicating the transition of

TABLE I. Simulation parameters of PEO solutions [a].

| $C$ | $R_e$ | $\rho$ | $C^*$ | $M$ | $\tau_0$ | $\tau_z$ |
|---|---|---|---|---|---|---|
| (%) | (nm) | (g cm$^{-3}$) | (%) | (g mol$^{-1}$) | ($10^5$ fs) | ($10^5$ fs) |
| 20 | 1.49 | 1.062 | 22.25 | 470.588 | 0.23 | 1.47 |
| 16 | 1.51 | 1.057 | 21.60 | 470.588 | 0.23 | 1.47 |
| 12 | 1.47 | 1.053 | 23.44 | 470.588 | 0.15 | 0.98 |
| 10 | 1.48 | 1.051 | 22.94 | 470.588 | 0.23 | 1.47 |
| 8 | 1.44 | 1.048 | 25.02 | 470.588 | 0.23 | 1.47 |
| 4 | 1.44 | 1.042 | 25.00 | 470.588 | 0.23 | 1.47 |

[a] See Appendix A for $\tau_0$ and $\tau_z$.

polymer solution from dilute solution to semi-dilute solution, can be estimated as follows:

$$C^* \approx \frac{M}{R_e^3}. \tag{10}$$

As shown in Table I, our simulation model satisfies the condition of dilute solution as $C^* > C$.

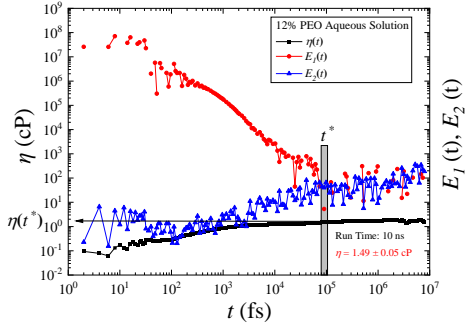

**FIG. 3.** Integration of shear stress modules and best estimations of cutoff time of 12% PEO solution. The curve of $\eta(t)$, $E_1(t)$, and $E_2(t)$, in which the results are averaged over six independent trajectories for 10 ns, are shown in black, red, and blue, respectively. $t^*$ is $1.47 \times 10^5$ fs and the estimated $\eta$ is $1.49 \pm 0.05$ cP. The best estimated $\eta$ of all test vehicles are obtained by following the same procedure, and the results are provided in Appendix B. $E_1(t)$ and $E_2(t)$ are defined by Eq. (8) and Eq. (9), respectively.

After obtaining the shear stress modules using the Green-Kubo formula [Eq. (3a)], the cutoff time $t^*$ was estimated by comparing the error of Eq. (8) to the error of Eq. (9). In practice simulations, the viscosity $\eta = \eta(t^*)$ can be obtained iteratively by the form by checking the ideal viscosity $\eta(\infty)$ to the simulation value $\eta(t^*)$ in Eq. (7b). Figure 3 shows the results of the PEO solution at 12% weight density.

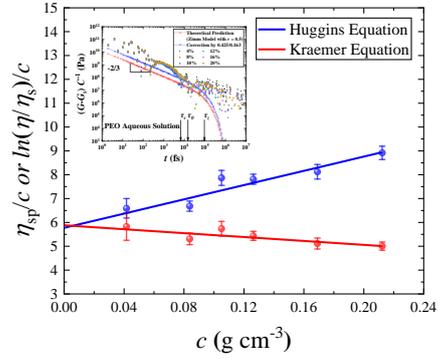

**FIG. 4.** Determining intrinsic viscosity of modeled PEO aqueous solutions. The red line represents the Kraemer equation, and the blue line the Huggins equation. The inset shows the comparison of simulations to the theoretical prediction of shear stress modulus (red and blue dotted lines) without adjusted parameters. The derivation of the correction coefficient is shown in Appendix A.

The algorithm could thus be verified through the Fox-Flory equation by using Huggins equation and Kraemer equation [30], as shown in Fig. 4. The intrinsic viscosity $[\eta]$ was approximately $5.99 \times 10^{-6}$ m$^3$ g$^{-1}$, and the constant in the Fox-Flory equation was approximately $8.4 \times 10^{23}$ mol$^{-1}$, which is in qualitative agreement with the theoretical prediction $\Phi = 0.425 N_{av} = 2.5 \times 10^{23}$ mol$^{-1}$. However, the deviation of the simulation result and the theoretical Flory-Fox constant $\Phi$ may be caused by the water viscosity error. Our simulation provides the viscosity of water to be approximately 0.75 cP where the experiment value is around 0.87 cP at 300K.

The relaxation time of the PEO chain in dilute solutions can be also estimated semi-quantitively by the cutoff time $\tau_Z \cong t^*$. The relaxation time of the Kuhn segment and PEO chain are summarized in Table I. The results are consistent with the prediction of Zimm dynamics of $\tau_Z \cong \tau_0 N^{3\nu}$, where $\nu = 0.5$ for $\Theta$ solvents. The shear stress modules of the PEO solutions and the theoretical prediction without adjusted parameters are shown in the inset of Fig 4. The simulation results are in good agreement with the best estimation of cutoff time.

Thus, our study results show that our algorithm can swiftly and steadily converge to $t^*$, which reflects the outstanding robustness of the procedure and the reproducibility of the final $\eta$. To further demonstrate the validity and trustworthiness of the viscosity calculation results, we attempt to compare these results with that of the theoretical model. Surprisingly, the decay curves of $G(\tau)$ exhibit qualitative agreement with Zimm Model (as shown in the inset of Fig 4), which has been regarded as the best model for polymer dynamics in a dilute solution [30]. However, because of chain length and range of concentration, the Zimm model may fail to predict the curve of time correlation function $G(\tau)$. This example demonstrates that our result is reliable, the intrinsic viscosity $[\eta]$ of the simulation system is predicted by Fox-Flory equation based on $\eta$, and the forecasted constant $\Phi$ is as per our expectation. Compared to all previous studies, particularly time decomposition method [13] and semiempirical analytic forms for time correlation functions [19-24], our criterion demonstrates: (1) simpler and more robust handling procedures without model-dependence; (2) more precise and reliable results in the fluctuation-based equilibrium simulation method.

**Conclusion** - In summary, we propose a universal approach for estimating the cutoff time of viscosity estimation using the Green-Kubo formula without any model dependence. The robustness of the algorithm is demonstrated by the simulation of viscosity of PEO aqueous solution, which is challenging in simulations with a long tail of correlations. The algorithm provides reliable results for the simulation system with multiple modes of correlation times and a relatively short run and can be applied in the simulations of complex systems with a long tail of correlations typically found in the macromolecular and biological simulation systems.

We acknowledge financial support from the National Natural Science Foundation of China under Grant 21574139, 21973103. The simulation was supported by the National Computing Tianjin Center.

## Appendix A: Derivation of Correction Coefficient

In the Zimm model, the monomer relaxation time $\tau_0$ and the chain relaxation time $\tau_z$ can be written in terms of the solvent viscosity $\eta_s$ [30],

$$\tau_0 = 0.163 \frac{\eta_s b^3}{k_B T} \qquad (A1)$$

$$\tau_z = 0.163 \frac{\eta_s}{k_B T} R_e^3 = \tau_0 N^{3\nu} \qquad (A2)$$

where $b$ is the Kuhn length, and $N = M/M_0$ is the number of Kuhn monomers of a chain (for PEO, $M_0 = 137 \text{ g mol}^{-1}$ [30] ); the relaxation time of the $p$th mode has a similar form to $\tau_z$

$$\tau_p \approx \tau_0 \left(\frac{N}{p}\right)^{3\nu} \quad \text{for } p = 1, 2, \dots, N . \qquad (A3)$$

For $t > \tau_0$, the $G(t)$ can be calculated as [31]

$$\begin{aligned} G(t) &= \frac{c_N}{N} k_B T \sum_N \exp\left(-\frac{t}{\tau_p}\right) \\ &= \frac{cRT}{M} \sum_N \exp\left(-\frac{t}{\tau_p}\right), \end{aligned} \qquad (A4)$$

where $c_N$ is the number of segments per volume, $c$ is the mass concentration, and $R$ is the molar gas constant. Due to Eq. (B3), an excellent approximation to $G(t)$ may be given as follows:

$$G(t) \approx \frac{cRT}{M} N \left(\frac{t}{t_0}\right)^{-\frac{1}{3\nu}} \exp\left(-\frac{t}{\tau_z}\right). \quad (A5)$$

Then, the polymer contribution to $\eta$ can be computed the Green-Kubo formula,

$$\eta - \eta_s = \int_0^\infty G(t)dt \quad (A6a)$$

$$\approx \frac{cRT}{M} N \int_{\tau_0}^\infty \left(\frac{t}{t_0}\right)^{-\frac{1}{3\nu}} \exp\left(-\frac{t}{\tau_z}\right) dt \quad (A6b)$$

$$\approx \frac{cRT}{M} N \tau_z \left(\frac{\tau_z}{t_0}\right)^{-\frac{1}{3\nu}} \int_{1/N^{3\nu}}^\infty (x)^{-\frac{1}{3\nu}} \exp(-x) \, dx \quad (A6c)$$

$$\approx 0.163 c\eta_s N_{Av} \frac{R_e^{\ 3}}{M} \int_{1/N^{3\nu}}^\infty (x)^{-\frac{1}{3\nu}} \exp(-x) \, dx \quad (A6d)$$

The transformation from Eq. (A6c) to Eq. (A6d) can be obtained using Eq. (A1) and Eq. (A2). By comparing our results [Eq. (A7c)] and the Fox-Flory equation [Eq. (A8)], we have

$$[\eta] = \lim_{c \to 0} \frac{\eta - \eta_s}{c\eta_s} \quad (A7a)$$

$$\approx 0.163 \int_{\frac{1}{N^{3\nu}}}^\infty (x)^{-\frac{1}{3\nu}} \exp(-x) \, dx \quad N_{Av} \frac{R_e^{\ 3}}{M} \quad (A7b)$$

$$\approx 0.163 N_{Av} \frac{R_e^{\ 3}}{M} \quad (A7c)$$

$$[\eta] = 0.425 N_{Av} \frac{R_e^{\ 3}}{M} \quad (A8)$$

we can obtain a correction coefficient for Eq. (A5), rewritten as

$$G(t) = \frac{0.425}{0.163} \frac{cRT}{M} N \left(\frac{t}{t_0}\right)^{-\frac{1}{3\nu}} \exp\left(-\frac{t}{\tau_z}\right) \quad (A9)$$

## Appendix B: Best Estimation of all Test Vehicles

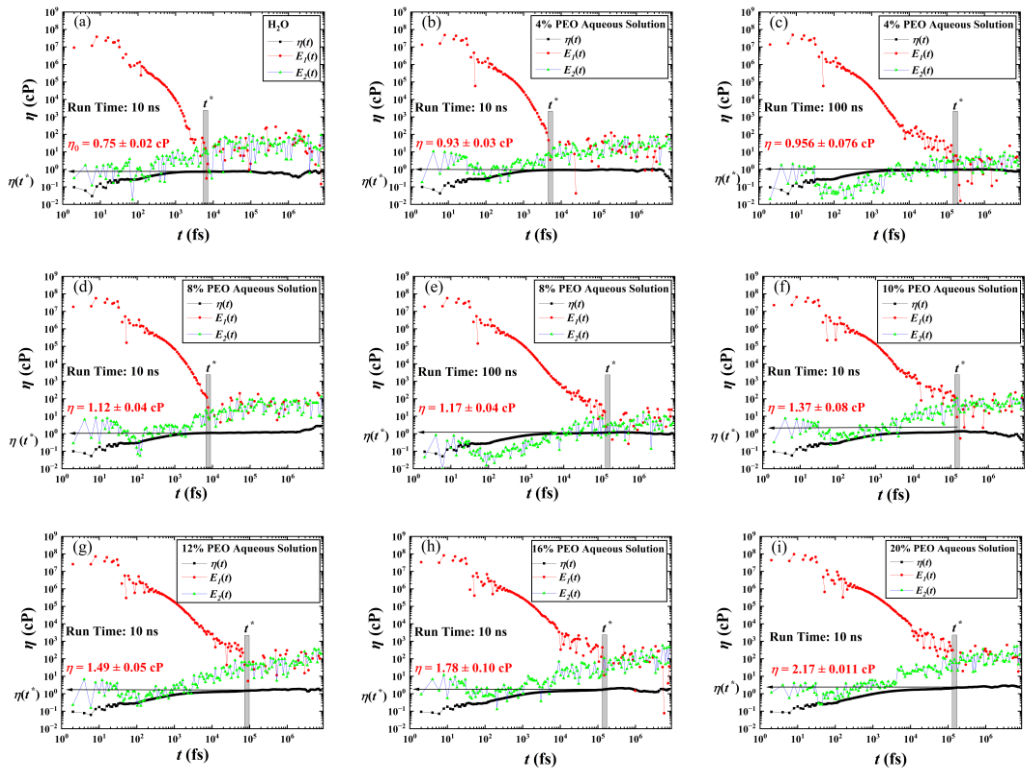

**FIG. B1.** Integration of shear stress modules and best estimations of cutoff time for all test vehicles.

* guoxuhong@ecust.edu.cn

† qiliao@iccas.ac.cn

[1] M. P. Allen and D. J. Tildesley, *Computer Simulation of Liquids* (Oxford University Press, New York, 2017).

[2] B. Hess, Determining the shear viscosity of model liquids from molecular dynamics simulations, J. Chem. Phys. **116**, 209 (2002).

[3] B. D. Todd and P. J. Daivis, *Nonequilibrium Molecular Dynamics: Theory, Algorithms and Applications* (Cambridge University Press, Cambridge, 2017).

[4] C. Hargus, K. Klymko, J. M. Epstein, and K. K. Mandadapu, Time reversal symmetry breaking and odd viscosity in active fluids: Green-Kubo and NEMD results, J. Chem. Phys. **152** (2020).

[5] T. Chen, B. Smit, and A. T. Bell, Are pressure fluctuation-based equilibrium methods really worse than nonequilibrium methods for calculating viscosities?, J. Chem. Phys. **131**, 246101 (2009).

[6] M. Schoen and C. Hoheisel, The shear viscosity of A Lennard-Jones fluid calculated by Equilibrium Molecular-Dynamics, Mol. Phys. **56**, 653 (1985).

[7] Y. Zhang, G. Guo, and G. Nie, A molecular dynamics study of bulk and shear viscosity of liquid iron using embedded-atom potential, Phys. Chem. Miner. **27**, 164 (2000).

[8] F. Taherkhani and H. Rezania, Temperature and size dependency of thermal conductivity of aluminum nanocluster, J. Nanopart. Res. **14**, 1222 (2012).

[9] J. W. Lawson, M. S. Daw, and C. W. Bauschlicher, Lattice thermal conductivity of ultra high temperature ceramics ZrB2 and HfB2 from atomistic simulations, J. Appl. Phys. **110**, 083507 (2011).

[10] G. S. Fanourgakis, J. S. Medina, and R. Prosmiti, Determining the bulk viscosity of rigid water models, J. Phys. Chem. A **116**, 2564 (2012).

[11] J. F. Danel, L. Kazandjian, and G. Zérah, Numerical convergence of the self-diffusion coefficient and viscosity obtained with Thomas-Fermi-Dirac molecular dynamics, Phys. Rev. E **85**, 066701 (2012).

[12] Z. Ren, A. S. Ivanova, D. Couchot-Vore, and S. Garrett-Roe, Ultrafast structure and dynamics in ionic liquids: 2D-IR spectroscopy probes the molecular origin of viscosity, J. Phys. Chem. Lett. **5**, 1541 (2014).

[13] Y. Zhang, A. Otani, and E. J. Maginn, Reliable viscosity calculation from equilibrium molecular dynamics simulations: a time decomposition method, J. Chem. Theory Comput. **11**, 3537 (2015).

[14] Y. Zhang, L. Xue, F. Khabaz, R. Doerfler, E. L. Quitevis, R. Khare, and E. J. Maginn, Molecular topology and local dynamics govern the viscosity of imidazolium-based ionic liquids, J. Phys. Chem. B **119**, 14934 (2015).

[15] O. A. Moultos, Y. Zhang, I. N. Tsimpanogiannis, I. G. Economou, and E. J. Maginn, System-size corrections for self-diffusion coefficients calculated from molecular dynamics simulations: The case of CO2, n-alkanes, and poly(ethylene glycol) dimethyl ethers, J. Chem. Phys. **145**, 074109 (2016).

[16] N. Kondratyuk, Contributions of force field interaction forms to Green-Kubo viscosity integral in n-alkane case, J. Chem. Phys. **151**, 074502 (2019).

[17] N. D. Kondratyuk and V. V. Pisarev, Calculation of viscosities of branched alkanes from 0.1 to 1000 MPa by molecular dynamics methods using COMPASS force field, Fluid Phase Equilib. **498**, 151 (2019).

[18] N. Wang, Y. Zhang, K. S. Al-Barghouti, R. Kore, A. M. Scurto, and E. J. Maginn, Structure and dynamics of Hydrofluorocarbon/Ionic liquid mixtures: an experimental and Molecular Dynamics study, J. Phys. Chem. B **126**, 8309 (2022).

[19] D. M. Heyes, E. R. Smith, and D. Dini, Shear stress relaxation and diffusion in simple liquids by molecular dynamics simulations: Analytic expressions and paths to viscosity, J. Chem. Phys. **150**, 174504 (2019).

[20] D. M. Heyes and D. Dini, Intrinsic viscuit probability distribution functions for transport coefficients of liquids and solids, J. Chem. Phys. **156**, 124501 (2022).

[21] P. Español, J. A. de la Torre, and D. Duque-Zumajo, Solution to the plateau problem in the Green-Kubo formula, Phys. Rev. E **99**, 022126 (2019).

[22] L. d. S. Oliveira and P. A. Greaney, Method to manage integration error in the Green-Kubo method, Phys. Rev. E **95**, 023308 (2017).

[23] D. Duque-Zumajo, J. A. de la Torre, and P. Espanol, Non-local viscosity from the Green-Kubo formula, J. Chem. Phys. **152**, 174108 (2020).

[24] P. Pattnaik, S. Raghunathan, T. Kalluri, P. Bhimalapuram, C. V. Jawahar, and U. D. Priyakumar, Machine learning for accurate force calculations in molecular dynamics simulations, J. Phys. Chem. A **124**, 6954 (2020).

[25] P. J. in 't Veld and G. C. Rutledge, Temperature-dependent elasticity of a semicrystalline interphase composed of freely rotating chains, Macromolecules **36**, 7358 (2003).

[26] S. Hezaveh, S. Samanta, G. Milano, and D. Roccatano, Molecular dynamics simulation study of solvent effects on conformation and dynamics of polyethylene oxide and polypropylene oxide chains in water and in common organic solvents, J. Chem. Phys. **136**, 124901 (2012).

[27] K.-S. Kim, M. H. Han, C. Kim, Z. Li, G. E. Karniadakis, and E. K. Lee, Nature of intrinsic uncertainties in equilibrium molecular dynamics estimation of shear viscosity for simple and complex fluids, J. Chem. Phys. **149**, 044510 (2018).

[28] A. P. Thompson *et al.*, LAMMPS-a flexible simulation tool for particle-based materials modeling at the atomic, meso, and continuum scales, Comput. Phys. Commun. **271**, 108171 (2022).

[29] R. W. Hockney and J. W. Eastwood, *Computer simulation using particles* (Taylor & Francis, Inc., Abingdon, 1988).

[30] M. Rubinstain, R. H. Colby, *Polymer Physics* (Oxford University Press, Oxford, 2003).

[31] M. Doi, S. F. Edwards, *The Theory of Polymer Dynamics* (Oxford University Press, Oxford, 1986).